\documentclass[twocolumn]{elsart}
\usepackage{natbib}
\begin{document}
\runauthor{Di\thinspace Stefano, Kong and Primini}
\begin{frontmatter}
\title{Ten Facts of Life for Distant Supersoft Sources}
\author[CfA]{Rosanne Di\thinspace Stefano} 
\author[MIT]{Albert Kong}
\author[CfA]{Francis A. Primini}
\thanks[1]{RD is also at Tufts University, Medford, MA 02155, USA}

\address[CfA]{Harvard-Smithsonian Center for Astrophysics, 
60 Garden Street, Cambridge, MA 02138, USA} 
\address[MIT]{Kavli Institute, MIT, Cambridge, MA 02139, USA}
\begin{abstract}
First discovered in the Magellanic Clouds and in the Milky Way, the largest pools of luminous supersoft X-ray sources (SSSs) now known lie in M31 and in more distant galaxies. Hundreds of newly-discovered SSSs are helping us to test models for Type~Ia supernovae and to identify SSSs that may represent a wider range of physical systems, including accreting intermediate-mass black holes.  In this short report we list ten intriguing facts about distant SSSs.     
\end{abstract}
\begin{keyword}
X-ray sources; supersoft sources: galaxies; 
Type~Ia supernova; symbiotic binaries, supernova remnants; 
intermediate-mass black holes 
\end{keyword}
\end{frontmatter}

\section{Introduction}
\vspace{-.3 true in} 
Luminous supersoft X-ray sources (SSSs)
were discovered and defined
in terms of
properties observed in a small number of sources ($\sim 18$)
 in the Galaxy
and Magellanic Clouds (MCs). Specifically, SSSs are
 defined in terms of their estimated luminosities ($L > 10^{36}$ erg s$^{-1}$)
and their broad band spectra, with little or no emission above
$1$ keV.  The physical nature of SSSs is not yet determined.
In fact, the studies we
summarize below indicate that they are likely to be a diverse group, with
several different types of physical systems observable as SSSs, including
white dwarfs (WDs), neutron stars (NSs), and black holes (BHs).
Perhaps this diversity is
to be expected, given the simplicity of the  definition.

\vspace{-.1 true in} 
SSS effective radii are comparable to those of
WDs. Indeed, roughly half of the SSSs in the MCs and
Milky Way with optical IDs
have counterparts that are consistent with systems known to contain
hot WDs: planetary nebulae, recent novae, and symbiotic 
binaries (Greiner 2000).
The remaining SSSs in
the Galaxy and MCs 
have measured periods ranging from a few hours to a few days.
The first and best-known model for these binary  was developed by 
Ed van den Heuvel and collaborators (1992).
The close-binary supersoft (CBSS) model, 
postulates that the prodigious luminosities are generated 
through the nuclear burning of material accreted by a WD. 
In order for nuclear burning to occur, the accretion rates must be high
(close to or larger than
$10^{-7} M_\odot$  year$^{-1}$).
These high rates can be sustained only if the donor star is 
somewhat more massive than the WD and/or
slightly evolved. 
An interesting characteristic of these models is that they allow
the WD to increase its mass.
Some SSS binaries may therefore be progenitors of Type~Ia
supernovae (Rappaport, Di~Stefano, \& Smith 1994; Di~Stefano 1996; 
Kahabka \& van den Heuvel 1997 and references therein).   

While indirect evidence mounts that the WD model
applies to some SSSs in the Galaxy and 
MCs, definite confirmation has been difficult.
(See the contribution of Phil Charles to this volume.)
One way to learn more about SSSs is to find them in other galaxies.
We expect
$\sim 1000$ 
SSSs to reside in spiral galaxies such as M31. (See  
Di~Stefano \& Rappaport 1994, Rappaport, Di~Stefano, \& Smith 1994,
Yungelson et al. 1996.)  
Thanks to {\it Chandra} and {\it XMM-Newton} 
we have begun to study SSSs in distant galaxies. Below we list some of the
things we have learned. 

\section{Ten Facts of Life}

{\bf 1. There is an extension of the class of SSSs to sources of
   somewhat higher energies.} 
If SSSs are nuclear-burning WDs, the hottest systems would be those
with masses near the Chandrasekhar limit, $M_C$; these could be
as hot as $150$ eV, with luminosities a few times $10^{38}$ erg s$^{-1}$.
Particularly if such a hot WD lies behind a large
absorbing column, its spectrum will appear to be harder than the spectra
of most SSSs. Because of their potential physical
importance, we must include any such sources
when selecting SSSs from among the X-ray sources detected in 
external galaxies. By creating a selection procedure sensitive to 
hot SSSs, we discovered that there are sources with
luminosities comparable to those of SSSs ($> 10^{36}$ erg s$^{-1}$),
but with effective temperatures that can be significantly higher,
generally between $100$ eV and $350$ eV. 
We refer to these somewhat hotter sources as quasisoft sources (QSSs).
Several distant QSSs are bright enough to allow spectral fits; many
of these are genuinely harder than the SSSs of the Galaxy and MCs.
Many are variable, and are therefore more likely to be X-ray binaries
than supernova remnants (SNRs). Although QSSs are almost certain to inhabit
our own Galaxy and the Magellanic Clouds, this class
has not yet been studied locally, and we do not know what they are.
To summarize,
we reserve the monicker SSS for sources with the characteristics of the
``classical" SSSs (Greiner 2000), while QSSs are harder, but have little or no
emission above $2$ keV.
We refer to any source that is either an SSS or a QSS as a {\it very
soft source}, a VSS.

{\bf 2. SSSs and QSSs are found in the bulges of spiral galaxies.} 
High Galactic absorption prevents us from detecting
SSSs in the Bulge of the Milky Way. 
{\it Chandra}, however, 
has discovered eight SSSs in the bulge of M31 that would satisfy the
standard criteria used to select SSSs (little or no emission above 
roughly $1$ keV). There are also several QSSs. 
(See Di~Stefano et al. 2004, and Orio 2005 for additional recent results from 
{\it XMM-Newton.})  
SSSs and QSSs 
   are also found in the bulges of other
spiral galaxies (Di~Stefano \& Kong 2003, 2004, and Di~Stefano  et al.\, 2003).

{\bf 3. VSSs are found very close to 
   the nuclei of both Local Group galaxies that house
supermassive black holes.}

An SSS lies within $2''$  of the central BH in M32 (Ho et al.\, 2003).
A QSS lies within $2''$  of the central BH in M31 (Garcia et al.\, 2000).
Given the local spatial densities of VSSs in these galaxies, the
probability that the projected position of a soft source is within
a few parsecs of the nucleus 
is very low. It therefore
seems likely that these sources are somehow related
to the presence of the supermassive BHs.
 A natural explanation is that the
sources are the hot cores of giants that were tidally stripped
by the massive central BH
(Di~Stefano et al.\, 2001). In fact, if tidal disruptions
of stars near supermassive BHs do occur, they should give rise to
bright flare events. The hot cores of stripped giants would be
the necessary complements. 
There are other possibilities. For example,  
the centrally located
VSSs may be interacting binaries 
that were formed via stellar interactions in the high-density
galactic bulge. Alternatively, the VSSs could contain intermediate-mass
black holes (IMBHs) that have migrated toward the galaxy's center.

{\bf 4. SSSs are found in elliptical galaxies and in globular clusters.}  
{\it Chandra} observations have discovered SSSs in elliptical galaxies
(See Table 1; Sarazin, Irwin, \& Bregman 2001; Di~Stefano \& Kong 2003, 2004),
and in the GCs of NGC 4472 and M104.  
The evolutionary scenario for CBSSs suggest that SSS binaries should be
present in elliptical galaxies. Novae and symbiotics are
also expected to be found in old stellar populations.

\begin{table*}
\caption{VSSs in galaxies}
\footnotesize
\begin{tabular}{lccccccccc}

\hline
\hline
Name   & Type& {\it Chandra} & Total & VSS & SSS & QSS & Canonical \\
       &     & exposure (ks) &       &     &     &     &\\ 
\hline
M101   &  Sc & 94.4          & 118   & 53  &32   & 21  & 65\\
M83    &  Sc & 49.5          & 128   & 54  & 28  & 26  & 74\\
M51    &  Sc & 28.6          & 92    & 36  & 15  & 21  & 56\\
M104   &  Sa & 18.5          & 122   & 22  & 5   & 17  & 100\\
NGC 4697 & E & 39.3          & 91    & 19  & 4   &15   & 72\\
NGC 4472 & E & 34.4          & 211   & 27  & 5   & 22  & 184\\

\hline

\end{tabular}
\par
\medskip
\begin{minipage}{0.95\linewidth}
\footnotesize
{\small
Notes---Both SSSs and QSSs have been found in every galaxy studied to date.
The examples in this table illustrate that VSSs constitute a smaller
fraction of all X-ray sources in galaxies dominated by older stellar
populations. Furthermore, in older populations the fraction of 
VSSs that are SSSs also
tends to be smaller.}  
\\

\end{minipage}
\par

\end{table*}

{\bf 5. A large fraction of SSSs in spiral galaxies are
associated with the spiral arms.} This suggests
that many SSSs in spirals are younger than
roughly $10^8$ years old. It seems unlikely that the 
majority of these systems are examples of close-binary supersoft
sources,  
because the typical
donor in a CBSS is a slightly evolved star with a mass of $1-2.5\, M_\odot.$
Such stars are most likely to be found away from spiral arms and regions
of star formation, because their ages are
greater than $10^8-10^9$ years. SSSs in 
the spiral arms could therefore
 be examples of different mass transfer scenarios, or
they could be binaries with neutron star or black hole accretors.

{\bf 6. Two of the most luminous X-ray sources ever discovered are
   SSSs. These may be good candidates for accreting intermediate-mass
black holes (IMBH).} In its high soft state, M101 ULX-1 is an SSS,
with luminosity that has been observed to reach $\sim 10^{41}$ erg s$^{-1}$. 
It has also been observed as a QSS and, in its lowest observed state, as a 
hard X-ray source (Kong \& Di~Stefano 2005). 
The high-state luminosity is incompatible with the WD
model, and may indicate that the accretor is a black hole
with mass larger than $10\, M_\odot.$ In fact, if the soft component
of the X-ray spectrum is emitted by the inner region of a   
multicolor disk, the inner disk radius is compatible with  $\sim 1000\, M_\odot$
BH. The state changes are also compatible with a BH accretor. 
(See also Fabbiano et al.\, 2003.)

{\bf 7. Most SSSs are highly time variable.}  
Time variability is a common characteristic of X-ray binaries.
SSSs are no exception. Many are transient and some have been observed
to vary within a single observation (see, e.g., {Swartz et al.(2002)}). 
The number of transients and the observed on/off patterns 
are inconsistent with the hypothesis that the transients
are dominated by recent novae. 

{\bf 8. Some SSSs are supernova remnants (SNRs).}  
In M31 contains an X-ray resolved SSS
with a counterpart detected in a narrow-band H$_\alpha$ image
(Kong et al.\, 2003).
Such a system raises the question of whether the SSSs we find in
spiral arms may also be SNRs. This question can be resolved by 
taking multiple images. In fact, as discussed above, SSSs tend
to be highly time variable, while SNRs cannot vary significantly over
short times. Although a definitive study has not yet been carried out,
the prevalence of variability, and the small fraction of
SNRs in the Local Group that are SSSs conspire to make it
seem unlikely that most spiral-arm SSSs are SNRs.

{\bf 9. Link to Type~Ia Supernovae:\ }
If an accreting WD is to experience a Type~Ia explosion, it must
be able to retain the matter it accretes. The nuclear burning of
accreted matter makes such retention possible, and this is the reason
that SSS-like behavior is expected for accreting WD (single degenerate)
 progenitors of Type~Ia supernovae.
In spiral galaxies, the rate of SNe~Ia is roughly $0.3$
per century per $10^{10} L_\odot$ in blue luminosity. If
a typical progenitor WD must gain roughly $0.2\, M_\odot$ in
order for an explosion occur, and if the rate of
mass accretion is $4 \times 10^{-7} M_\odot$ yr$^{-1}$,
then a galaxy such as M31 or M101 should contain roughly
$1000$ SSSs that are Type~Ia progenitors.
This is roughly
consistent with previous estimates of the total population of SSSs
in spiral galaxies.
There is a problem, however, since the accreting WDs which
are Type~Ia progenitors shold tend to have the highest WD
masses among the accretors, and to therefore be among the
brightest, hottest, and most detectable SSSs. In fact,
we can already rule out the possibility that most Type~Ia
progenitors in M31 and M101 are detectable as SSSs. This
is so both for models in which the WD needs to
achieve the Chandrasekhar mass in order to explode and for
a range of sub-Chandrasekhar models. (See Di\thinspace Stefano 2006 for
details.) This result does not eliminate single degenerate
models from consideration as possible Type~Ia progenitors.
It does, however, indicate that, if a significant fraction of
Type~Ia progenitors are accreting WDs, then they must be
internally obscured. In fact, this is consistent with
initial calculations for Type~Ia progenitor models, and may be an
indication that high rates  of accretion are, as the calculations
indicate, often accompanied by high rates of mass ejection.
The ejected mass can then absorb soft radiation from the system.

{\bf 10. QSS Models: } Just as the type of X-ray emission that defines SSSs
is associated with different types of physical
systems, QSS-like behavior is likely to be exhibited
by a variety of systems. Some QSSs may be SNRs, but variable QSSs
are more likely to be X-ray binaries. One of the M31 QSSs which happens to
be only slightly harder than a typical SSS has been tentatively identified
as a symbiotic (Di~Stefano et al.\, 2004b). Just as QSS broadband spectra
are extensions of SSS spectra from $1$ to $2$ keV, 
it seems likely that QSS physical systems
may represent extensions of those observed as SSSs.
It could be, e.g., that some of the soft emission is
reprocessed, with a fraction of it re-emitted at slightly higher
energies.  

\vspace{-.3 true in}

\section{Summary}
\vspace{-.3 true in} 

When Ed and his collaborators carried out the first
theoretical investigations of the newly established
class of SSSs, it was clear that something
exciting was going on. Fifteen years later, the
 study of SSSs in other galaxies
is leading to new and unexpected discoveries and challenges.
As we begin to develop a statistical sample
of SSSs, we are finding that they are a diverse group,
and that there are other soft sources, QSSs,
 that may be related to them. The realms of astrophysics
associated with SSSs and QSSs span supermassive BHs,
intermediate-mass BHs, X-ray binaries with stellar-remnant
accretors, Type~Ia progenitors, and supernova remnants.
If, fifteen years hence, we have succeeded in 
establishing the nature of the sources we detect, and
developing theoretical constructs to describe them and the roles 
they play in the universe, we will
have accomplished a good deal.  

\vspace{-.3 true in} 
{}

\end{document}